\documentclass[11pt,paper]{article}
\usepackage{jheppub}
\usepackage[usenames,dvipsnames,table]{xcolor}

\usepackage{soul}
\usepackage{caption}

\title{Stochastic gravity and turbulence}

\author[a]{Sebastian Waeber,} 
\author[a]{and Amos Yarom}
\affiliation[a]{Department of Physics, Technion, Haifa 32000, Israel}

\emailAdd{wsebastian@campus.technion.ac.il, ayarom@physics.technion.ac.il}

\begin{document}

\abstract{
We study the ensemble average of the thermal expectation value of an energy momentum tensor in the presence of a random external metric. In a holographic setup this quantity can be read off of the near boundary behavior of the metric in a stochastic theory of gravity. By numerically solving the associated Einstein equations and mapping the result to the dual boundary theory, we find that the non relativistic energy power spectrum exhibits a power law behavior as expected by the theory of Kolmogorov.
}

\maketitle

\section{Introduction}
Turbulence is a ubiquitous phenomenon which manifests itself whenever fluid flow at high Reynolds number is the appropriate physical kinematic description of the system, be it the quark gluon plasma generated in heavy ion collisions, fast flowing rivers, or neutron star evolution. Unfortunately, while it is a robust phenomenon, there is much we do not know regarding turbulent behavior. The most prominent characteristic of turbulent flow is the power law behavior of velocity field correlators. Kolmogorov's theory of turbulence \cite{K41} correctly predicts the power law behavior of the lower moments of these correlators. The behavior of higher moments is currently inexplicable.

At a what seems like a completely different corner of physical theories, lies the gauge-gravity duality \cite{Maldacena:1997re,Witten:1998qj,Gubser:1998bc} inspired by string theory. This duality relates, in an appropriate limit, gravitational dynamics in asymptotically anti de Sitter (AdS) space to that of conformal field theory. Since conformal field theories can be described by hydrodynamics, in its regime of validity, then by the duality, this hydrodynamic behavior must be encoded in gravitational dynamics of asymptotically AdS black branes.

If gravity, and black holes in particular, capture turbulent dynamics then this opens a wide range of possibilities for understanding the latter in terms of the former. In particular, universal power law behavior of moments of the velocity field should be captured by universal black hole behavior. While it stands to reason that the duality implies such a relation, the chief property of black holes  responsible for turbulent flow is somewhat of a mystery. Various attempts at identifying these key features of black holes can be found in the literature. 

Of particular relevance to our current study is the work of \cite{Bhattacharyya:2008jc} which made the connection between the constitutive relations of fluid dynamics and gravity in the presence of a negative cosmological constant manifest. This work allowed the authors of \cite{Adams:2013vsa} (see also \cite{Eling:2010vr,Carrasco:2012nf}) to obtain a conjectural relation between the power law behavior of the energy power spectrum, predicted by Kolmogorov, and a certain fractal dimension of the  apparent horizon. The authors then demonstrated that this relation holds in the case where the apparent horizon was initially excited and then left to decay---corresponding to decaying turbulent flow. 
In recent years there has been a significant body of work scrutinizing, and building on, the results of 
\cite{Adams:2013vsa}, see, e.g., \cite{Green:2013zba,Balasubramanian:2013yqa,Eling:2013sna,Eling:2015mxa,Westernacher-Schneider:2015gfa,Westernacher-Schneider:2017snn,Rozali:2017bll,Westernacher-Schneider:2017xie,Eyink:2017zfz,Chen:2018nbh,Andrade:2019rpn}. Arguably, one of the main shortcomings of \cite{Adams:2013vsa} is that it deals with decaying turbulence where the Kolmogorov scaling relations are notoriously difficult to capture for a prolonged time. Indeed, steady state turbulent flow is usually achieved by introducing a random driving force which continuously pumps energy into the fluid. In the context of the gauge-gravity duality such a driving force should be reflected by a random source for the  bulk fields. Put differently, a random driving force is dual to random boundary conditions for the dynamical fields, including the metric itself. While some of the aforementioned works (\cite{Balasubramanian:2013yqa,Westernacher-Schneider:2015gfa,Westernacher-Schneider:2017snn,Westernacher-Schneider:2017xie,Chen:2018nbh,Andrade:2019rpn} in particular, and also \cite{Bhattacharyya:2008ji}) have dealt with driving forces, there is no work, to our knowledge, which properly deals with a dual gravitational setup with random boundary conditions for the metric. The current work aims to fill in this gap. 

We have found that the main challenge in incorporating a random boundary metric in an asymptotically AdS geometry is to obtain a well defined set of stochastic differential equations. The crux being the non linear nature of the Einstein equations. Not every type of random source will yield well defined equations. In particular, as we argue in section \ref{S:stochastic}, a boundary metric characterized by white noise is incompatible with the gravitational dynamics at hand. Instead of white noise, we have found that the metric must be
characterized by an autocorrelation function which is not a distribution and is differentiable at least up to an appropriate, finite, order in derivatives.

After incorporating a stochastic boundary metric in a holographic setup, we solved the stochastic Einstein equations numerically using a modest grid, sufficient to exhibit Kolmogorov scaling behavior of the dual (non relativistic) energy power spectrum across a decade of momenta  (see figure \ref{F:powerspectrum}). We found, among other things, that the near AdS boundary behavior of the metric (which determines the behavior of the stress tensor in the dual theory)  is enhanced in the deep interior due to the AdS warp factor. Thus, the horizon provides an amplified picture of the dynamics of the fluid---inline with qualitative notions of holography. We discuss our results in section \ref{S:discussion}.

\section{The metric as a stochastic driving force}
\label{S:stochastic}

We often use the generating function
\begin{equation}
	Z[g_{\mu\nu}] = \int D\phi e^{-S[\phi;\,g_{\mu\nu}]}
\end{equation}
to characterize correlation functions of the energy momentum tensor in a theory with some dynamical variables $\phi$ and an external metric $g_{\alpha\beta}$, e.g.,
\begin{equation}
	\langle 0 | T^{\mu \nu} | 0 \rangle{}_{g} = \frac{2 i}{\sqrt{g}}\frac{\delta}{\delta g_{\mu\nu}} Z[g_{\mu\nu}]\,.
\end{equation}
Here, 
and in the rest of this work, $\mu,\nu = 0,\ldots d-1$  will denote the spacetime directions of the field theory. The ensemble averaged vacuum expectation value of a stress tensor in a quantum field theory whose metric is characterized by a random variable, $Q_{\alpha\beta}$, 
\begin{equation}
\label{E:randomg}
	g_{\alpha\beta} = \eta_{\alpha\beta} + Q_{\alpha\beta}\,,
\end{equation}
is given by
\begin{equation}
	\overline{\langle 0 | T^{\mu\nu} | 0 \rangle} = \int D Q_{\alpha\beta} P(Q_{\alpha\beta}) \langle 0 | T^{\mu \nu} | 0 \rangle{}_{\eta+Q}\,,
\end{equation}
where $P(Q)$ specifies the distribution from which $Q_{\alpha\beta}$ is drawn. We will refer to $Q_{\alpha\beta}$ as the random, or stochastic, component of the metric. (See \cite{Aharony:2018mjm} for a study of renormalization group flow in a related setup.)

In this work we will be interested in thermal expectation values and not vacuum expectation values of the energy momentum tensor in the presence of the metric \eqref{E:randomg}. However, before proceeding, it is useful to pause and consider more broadly the properties of $T^{\mu\nu}$ in the presence of \eqref{E:randomg}. Coordinate invariance implies the Ward identity 
\begin{equation} 
\label{E:conservation}
	\nabla_{\mu}  T^{\mu\nu} = 0\,.
\end{equation}
The stochastic component of the metric, $Q_{\alpha\beta}$, contributes to \eqref{E:conservation} through the Christoffel connection and through a possible explicit dependence of the energy momentum tensor on the external metric \eqref{E:randomg}. Thus, at the very least, \eqref{E:conservation} will be a stochastic differential equation (SDE) where the external random variable ($Q_{\alpha\beta}$ in our case) is differentiated and may also appear non linearly. 
We refer the reader unfamiliar with the basics of stochastic differential equations to appendix \ref{A:review}, or to the extensive literature on the subject specified therein.

Stochastic differential equations of the form \eqref{E:conservation} are not well defined if the stochastic variable is characterized as white noise. Indeed, suppose $\xi(t,\vec{x})$ denotes white noise with zero average, i.e., $\overline{\xi(t,\vec{x})}=0$ and $\overline{\xi(t,\vec{x})\xi(t',\vec{x}')}=D(x-x') \delta(t-t')$ where $D$ is a smooth function of its arguments and a bar denotes an ensemble average. It follows that rational functions of $\xi(t,\,\vec{x})$ are ill defined, as are functions of its temporal derivatives. 

To make sense of \eqref{E:conservation} in the presence of \eqref{E:randomg}, we use a random force which is determined via an Ornstein-Uhlenbeck process, or, in more involved cases, via what we refer to as a cascade of Ornstein-Uhlenbeck processes. Recall that, in a $0+1$ dimensional setting, an Ornstein-Uhlenbeck process, $Q(t)$, is defined by
\begin{equation}
	\dot{Q}(t) = -\frac{Q(t)}{\tau_1} + \frac{\xi(t)}{\tau_2}
\end{equation}
where a dot denotes a time derivative, $\tau_1$ and $\tau_2$ are positive numbers, and $\xi$ denotes white noise with zero average. In what follows we use, for simplicity, $\tau_1 = \tau_2 = \tau$. We define a cascade of Ornstein-Uhlenbeck processes of order $n$, $Q(t)$, $q_1(t)$, $\ldots$, $q_n(t)$ via
\begin{equation}
\label{E:cascade}
	\dot{Q}(t) = -\frac{Q(t)}{\tau} + \frac{q_1(t)}{\tau}\,,
	\quad
	\ldots\,,
	\quad
	\dot{q}_i(\tau) = -\frac{q_i(t)}{\tau} + \frac{q_{i-1}(t)}{\tau} \,,
	\quad
	\ldots\,,
	\quad
	\dot{q}_n(\tau) = -\frac{q_n(t)}{\tau} + \frac{\xi}{\tau}\,,
\end{equation}
where, as before, $\xi$ denotes white noise with zero mean.
(So that we may refer to an Ornstein-Uhlenbeck process as a cascade of order $0$.) 
Using \eqref{E:cascade} it is straghtforward to argue that an SDE of the form
\begin{equation}
\label{E:nonlinearSDE}
	\dot{X}(t) = G\left(X(t),\,Q(t),\,\dot{Q}(t),\,\ddot{Q}(t),\,\ldots,\left(\frac{\partial}{\partial t}\right)^n Q(t) \right)
\end{equation}
can be rewritten as a set of (well defined) coupled stochastic differential equations linear in the white noise $\xi(t)$ (see e.g., \eqref{E:coupledIto}) as long as $Q$, $q_1$, $\ldots$, $q_n$, are given by a cascade of at least order $n$. If the $n$'th derivative in \eqref{E:nonlinearSDE} appears linearly then a cascade of order $n-1$ will suffice.

In order to completely specify the problem it remains to state the integration scheme associated with the stochastic noise term: as explained in appendix \ref{A:review}, apart from initial conditions, one needs to specify an integration scheme for an SDE in order to obtain a unique solution to it. Which integration scheme to use depends on the problem at hand. Often the It\^{o} integration scheme is used when studying formal aspects of SDE's while the Stratonovich scheme is more common in physical problems. In the current case, the stochastic metric is external so that the associated integration scheme may be determined at our discretion. After some experimentation we have decided on using the Stratonovich scheme. Using an It\^{o} scheme often resulted in what seemed like unphysical features. In particular, we observed an overall decrease in the energy density over time. 

While equations \eqref{E:cascade} and \eqref{E:nonlinearSDE} have been specified in a $0+1$ dimensional setting, they can be easily generalized to include a spatial dependence. In the remainder of this work we will focus on a $2+1$ dimensional spacetime, where the Cartesian spatial coordinates have period $L$, $x^i \sim x^i+L$. This implies, among other things, that momentum is discrete, $\vec{k}_i = \frac{2\pi \vec{n}_i}{L}$ with $\vec{n}_i$ a pair of integers. We choose a random metric 
\begin{subequations}
\label{E:StochasticF}
\begin{equation}
\label{E:Fcomponents}
	Q_{\alpha,\beta}(t,\vec{x}) = \left(e^{2Q(t,\vec{x})}-1\right) \sum_{i=1,2}\delta_{\alpha\,x^i}\delta_{\beta\,x^i}
\end{equation}
such that
\begin{equation}
\label{E:Qtoq}
	e^{2Q(t,\vec{x})}-1 = q(t,\vec{x}) + 3 \sqrt{ \frac{D}{2 \tau} N_A \left(1 - e^{-\frac{2t}{\tau}}\right) }\,.
\end{equation}
The first term on the right hand side of \eqref{E:Qtoq}, $q$, is determined by a cascade of order $0$, 
\begin{equation}
\label{E:dotq}
	\dot{q}(t,\vec{x}) = - \frac{q(t,\vec{x})}{\tau} + \frac{\xi(t,\vec{x})}{\tau}\,,
\end{equation}
with $\xi(t,\vec{x})$ a random variable satisfying $\overline{\xi(t,\vec{x})}=0$ and
\begin{equation}
\label{E:averagexi2}
	\overline{\xi(t,\vec{x})\xi(t',\vec{x}')} = D \delta(t-t') \sum_{{i ,\, \vec{k}_i \in A}} \cos\left(\vec{k}_i\cdot \left(\vec{x}-\vec{x}'\right) \right)\,,
\end{equation}
\end{subequations}
the sum running over all momenta $\vec{k}_i$ inside an annulus $A$ of inner radius $k_f-\Delta k$ and outer radius $k_f+\Delta k$. The parameter $D$ which characterizes the strength of the random force is positive. In the second term on the right hand side of \eqref{E:Qtoq} we have defined 
\begin{equation}
	N_A = \sum_{i,\,\vec{k}_i \in A} 1\,, 
\end{equation}
the `size' of the annulus.

In place of \eqref{E:Qtoq} and \eqref{E:dotq}, we could have chosen $Q$ itself to be specified by a cascade of order $0$ which would have resulted in a skewed distribution around $e^Q=1$. Instead, we have chosen to parameterize $e^{Q}$ as in \eqref{E:Qtoq} where the second term on the right ensures that the left hand side is (statistically speaking) positive;
\begin{equation}
	\overline{q(t,\vec{x})q(t,\vec{x})} = \frac{D}{2 \tau} N_A \left(1 - e^{-\frac{2t}{\tau}}\right)\,.
\end{equation}
Correlations of the form \eqref{E:averagexi2} can be obtained from the distribution 
\begin{equation}
\label{E:xidistribution}
	\xi(t,\vec{x}) = \sqrt{D}\sum_{i,\,\vec{k}_i \in A} \left( P^{(1)}_{i}(t) \cos \left(  \vec{k}_i \cdot \vec{x} \right)+P^{(2)}_{i}(t)\sin \left( \vec{k}_i \cdot \vec{x} \right)\right)
\end{equation}
where the $P^{(a)}_{i}(t)$ are independent random variables. At each time $t$ and for each $a=1,2$ and $\vec{k}_i \in A$, $P^{(a)}_{i}(t)$ is drawn from a normal distribution, $N(0,1)$. 
In principle, we could have omitted the  sum over the annulus in \eqref{E:xidistribution} and obtained the same ensemble average as in \eqref{E:averagexi2}. The  sum in \eqref{E:xidistribution} ensures that each element of the ensemble will have an isotropic driving force. Note that spatial derivatives of $\xi$ are well defined---its stochastic component enters only through the $P_i^{(a)}$'s.

The metric specified by \eqref{E:randomg} and \eqref{E:StochasticF} allows us to treat conservation equations which are at most linear in time derivatives of the random metric but may contain arbitrary powers of it or its spatial derivatives. As we will see shortly, \eqref{E:Fcomponents} has been chosen such that the resulting equations of motion are precisely of this form.

\section{A holographic realization}
Our main goal in this work is to compute the average value of the thermal expectation value of the energy momentum tensor, in the presence of a stochastic boundary metric given by \eqref{E:randomg} and \eqref{E:StochasticF},
\begin{equation}
\label{E:defT}
	\mathcal{T}^{\mu\nu} =\overline{ \hbox{Tr} \left(\frac{e^{-\beta H}}{\hbox{Tr}(e^{-\beta H})} T^{\mu\nu}\right) }\,.
\end{equation}
We will do so using the holographic duality \cite{Maldacena:1997re} which relates the dynamics of field theories to the dynamics of asymptotically anti de Sitter (AdS) spaces. In particular, suppose
\begin{equation}
\label{E:solution}
	ds^2 = \frac{\ell^2}{\zeta^2} \left( \sum_{k=0}^{\infty} g_{\mu\nu}^{(k)}(\vec{x}) \zeta^{k}dx^{\mu} dx^{\nu} + d\zeta^2 \right)
\end{equation}
(with latin indices $m,n = 0,\ldots 3$ (in contrast to $\mu,\nu = 0,\ldots, 2$) which do not run over the `$\zeta$' coordinate) is a solution to the equations of motion associated with the Einstein-Hilbert action
\begin{equation}
\label{E:EHCaction}
	S = \frac{1}{16\pi G_N} \int d^4x \sqrt{g} \left( R + \frac{6}{\ell^2} \right) \,.
\end{equation}
This solution specifies a state, $\varrho$, in the dual field theory, such that the expectation value of the energy momentum tensor in this state is given by
\begin{equation}
\label{E:dictionary}
	\hbox{Tr} \left( \varrho {{T}_{\mu\nu}} \right)\Bigg|_{g_{\mu\nu} = g^{(0)}_{\mu\nu}} = \frac{3 \ell^2}{16 \pi G_N} g^{(3)}_{\mu\nu} \,,
\end{equation}
(so that the field theory is placed on a manifold with metric $g^{(0)}_{\mu\nu}$) where $\ell^2/G_N$ is often associated with a large number tied to the gauge group of the dual field theory. See, e..g, \cite{deHaro:2000vlm} for details.
 
For instance, the black brane geometry
\begin{equation}
\label{E:BBFG}
	ds^2 = \frac{\ell^2}{\rho^2} \left(-\left(1-\left(\frac{\rho}{\rho_0}\right)^3\right)dt^2 + dx_1^2+dx_2^2 - 2 dt d\rho  \right)
\end{equation}
is a solution to the Einstein equations following from \eqref{E:EHCaction} with Hawking temperature 
\begin{equation}
\label{E:THawking}
	T_0 = \frac{3}{4\pi \rho_0}\,.
\end{equation}
Bringing \eqref{E:BBFG} to the form \eqref{E:solution} leads, via \eqref{E:dictionary}, to 
\begin{equation}
	\hbox{Tr} \left(\frac{e^{-\beta H}}{\hbox{Tr}(e^{-\beta H})} T^{\mu\nu}\right) = \frac{3 \ell^2}{16\pi G_N} \frac{64\pi^3 T_0^3}{81}\begin{pmatrix} 2 & 0 & 0  \\ 0 & 1 & 0 \\ 0 & 0 & 1 \end{pmatrix}\,,
\end{equation}
the expected expression for the energy momentum tensor of a conformal field theory, in flat space, in a thermal state.

We propose that a holographic expression for $\mathcal{T}^{\mu\nu}$ defined in \eqref{E:defT} is given by
\begin{equation}
\label{E:averageT}
	\mathcal{T}^{\mu\nu} = \frac{3 \ell^2}{16 \pi G_N} \overline{g^{(3)\mu\nu}}
\end{equation}
where $g^{(3)\mu\nu} = g^{(0)\mu\alpha}g^{(0)\nu\beta}g^{(3)}_{\alpha\beta}$ and $g^{(0)\mu\nu}$ is the inverse of $g^{(0)}_{\mu\nu}$. The averaging in \eqref{E:averageT} is carried out over boundary metrics $g_{\mu\nu}^{(0)}$ specified in \eqref{E:randomg} which in the current notation reads
\begin{equation}
\label{E:Boundaryg}
	g^{(0)}_{\mu\nu}dx^{\mu}dx^{\nu} = -dt^2 + e^{2Q(t,\vec{x})}(dx_1^2 + dx_2^2)\,.
\end{equation}
We choose a $Q$ such that $Q=0$ for $t<0$ and is a stochastic variable, defined by \eqref{E:StochasticF}, at later times. Put differently, in order to obtain $\mathcal{T}^{\mu\nu}$ we must solve the Einstein equations derived from \eqref{E:EHCaction} with boundary conditions given by \eqref{E:Boundaryg} while treating the boundary value of the metric, and therefore the metric itself, as a random variable at times $t>0$.

To set up the problem it is convenient to use the infalling coordinate system suggested by \cite{Chesler:2013lia} such that
\begin{equation}
\label{E:EFcoordinates}
	\frac{1}{\ell^2} ds^2 =  \Sigma(t,\vec{x},\rho)^2 \hat{g}_{ij}(t,\vec{x},\rho) dx^{i}dx^{j} - 2 dt \left(F_{i}(t,\vec{x},\rho) dx^{i} +A(t,\vec{x},\rho)dt + \frac{d\rho}{\rho^2}\right) 
\end{equation}
where $|\hat{g}|=1$, $t = x^0$ is the time coordinate, and $i=1,2$ are the spatial coordinates. The radial coordinate $\rho$ vanishes at the asymptotic AdS boundary. The overall factor of $\ell$ in \eqref{E:EFcoordinates} has been conveniently chosen so that it drops out of the equations of motion:
\begin{subequations}
\label{E:EOM}
\begin{align}
\label{E:SigmaEOM}
	\left(\partial_{\rho}^2 + \frac{2}{\rho}\partial_{\rho} + \frac{1}{8}\, \hat g^{ij}\,(\partial_\rho \hat g_{jk})\, \hat g^{kl}\, (\partial_\rho \hat g_{li})  \right) \Sigma &= 0 \\
\label{E:FEOM}
	\left(\delta_i^j \partial_{\rho}^2 + \delta_i^j  \frac{2}{\rho}\partial_{\rho} + P_{F}{}_i^j [\hat{g},\,\Sigma] \right) F_j &= S_{F\,i}[\hat{g},\,\Sigma] \\
\label{E:d0SigmaEOM}
	\left(\partial_{\rho} + \frac{\partial_{\rho}\Sigma}{\Sigma}\right)d_0\Sigma & = S_{d_0\Sigma}[\hat{g},\,\Sigma,\,F]\\
\label{E:d0gEOM}
	\left( \delta^k_{(i}\delta^l_{j)}\partial_\rho +  \frac{\partial_\rho \Sigma}{\Sigma}\delta^k_{(i}\delta^l_{j)} -  \hat{g}^{mk}(\partial_\rho \hat{g}_{m(i} \delta^l_{j)})\right)d_0\hat{g}_{kl} & = S_{d_0\hat{g}\,}{}_{ij}[\hat{g},\,\Sigma,\,F,\,d_0\Sigma]\\
\label{E:AEOM}
	\left(\partial_{\rho}^2 + \frac{2}{\rho}\partial_{\rho}  \right) A &= S_{A}[\hat{g},\,\Sigma,\, F,\,d_0\Sigma,\,d_0\hat{g}]\\
\label{E:d0FEOM}
	\left(\delta^j_i \partial_{\rho}  -\frac{\partial_{\rho}\Sigma}{\Sigma} \delta^{j}{}_{i} - \hat{g}^{jk} \partial_{\rho} \hat{g}_{ki} \right) d_0 F_j &= S_{d_0 F}{}_{i}[\hat{g},\,\Sigma,\, F,\, d_0\Sigma,\, d_0\hat{g}, A] \\
\label{E:d0d0SigmaEOM}
	d_0d_0\Sigma & =  S_{d_0d_0\Sigma}[\hat{g},\,\Sigma,\, F,\, d_0\Sigma,\, d_0\hat{g},\,A,\,d_0F]
\end{align}
\end{subequations}
where we have defined
\begin{equation}
	d_0 = \partial_t - \rho^2 A \partial_{\rho}
\end{equation}
and 
\begin{align}
\begin{split}
	P_{F}{}^j_i[\hat{g},\,\Sigma] =&  -\hat{g}^{jk}\partial_{\rho}\hat{g}_{ki} + \left( 
		-\left(\partial_{\rho^2}\hat{g}_{ik} + \frac{2}{\rho} \partial_{\rho} \hat{g}_{ik} \right) \hat{g}^{kj} 
		+ \partial_{\rho}\hat{g}_{ik}\hat{g}^{kl}\partial_{\rho}\hat{g}_{lm}\hat{g}^{lj} - \frac{2\partial_{\rho}\Sigma}{\Sigma} \partial_{\rho}\hat{g}_{ik}\hat{g}^{kj}
		\right) \\
		&+\delta_i^j \left( \frac{1}{4}\, \hat g^{ml}\,(\partial_\rho \hat g_{lk})\, \hat g^{kp}\, (\partial_\rho \hat g_{pm}) - \frac{2\left(\partial_{\rho}\Sigma\right)^2}{\Sigma^2} \right)\,.
\end{split}
\end{align}
Expressions for the inhomogenous terms appearing on the right hand side of \eqref{E:EOM} can be found in \cite{Chesler:2013lia}.
The boundary conditions for the metric \eqref{E:Boundaryg} imply
 \begin{align}
\begin{split}
\label{E:BCs}
	\Sigma = \frac{e^{Q(t,\vec{x})}}{\rho} + \mathcal{O}(\rho^0)\,,
	\quad
	A = \frac{1}{2\rho^2} + \mathcal{O}(\rho^{-1})\,,
	\quad
	\hat{g}_{ij} = \delta_{ij} + \mathcal{O}(\rho)\,,
	\quad
	F = \mathcal{O}(\rho)\,,
\end{split}
\end{align}
and we also require that the metric is regular across the  apparent horizon. Our initial conditions are such that the metric takes the form \eqref{E:BBFG} at $t<0$ which ensures that the dual field theory is in thermal equilibrium at early times.

To relate the solution to \eqref{E:EOM} to the average value of the stress tensor \eqref{E:averageT}, consider the near boundary asymptotic expansion of the dynamical fields under the equations of motion,
\begin{align}
\begin{split}
\label{E:asymptotics}
	\Sigma =& \frac{e^{Q}}{\rho} + e^{Q} \lambda(t) + \mathcal{O}(\rho^3) \\
	\hat{g}_{ij} =& \delta_{ij} + \gamma_{ij} \rho^3 + \mathcal{O}(\rho^4) \\
	F_i =& f_i(t,\,\vec{x})  \rho + \mathcal{O}(\rho^2) \\
	A =& \frac{1}{2\rho^2} +\frac{ \lambda(t) - \partial_t{Q} }{\rho} 
	+ \left(\frac{\lambda^2(t)}{2} - \partial_t{\lambda}(t) - \lambda(t) \partial_t  Q  + \frac{1}{2} e^{-2 Q} \delta^{ij} \partial_i \partial_j Q  \right) 
	+ a(t,\,\vec{x}) \rho\,.
\end{split}
\end{align}
The undetermined parameter $\lambda(t)$ is a reflection of a residual coordinate transformation
\begin{equation}
\label{E:lambdatransformation}
	\rho \to \bar{\rho}(\rho) = \frac{\rho}{1+\rho \lambda(t)}\,,
\end{equation}
which preserves both the ansatz \eqref{E:EFcoordinates} and the boundary conditions \eqref{E:BCs}. In \cite{Chesler:2013lia} an extended version of this gauge freedom was used to tune the parameteric location of the apparent horizon to a convenient location. In what follows we will set $\lambda=0$.  

The functions $a$, $f_i$ and $\gamma_{ij}$ are determined by the near horizon boundary conditions. Together with $Q$, they specify the expectation value of the stress tensor. Using \eqref{E:dictionary} we find
\begin{align}
\begin{split}
\label{E:holographicTmn}
	\frac{16 \pi G_N}{3 L^2} \hbox{Tr}\left(\varrho T^{\mu\nu}\right) = & e^{-2Q} \begin{pmatrix}
		-\frac{4}{3} e^{2 Q} a &  f_1 & f_2 \\
		f_1 & -\gamma_{22} - \frac{2}{3} a &   \gamma_{12}  \\
		f_2 & \gamma_{12}  &  \gamma_{22} - \frac{2}{3}  a 
		\end{pmatrix} \\
		&-\frac{2}{3} e^{-2 Q} \delta^{0(\mu}\delta^{\nu)i}\partial_i\left(e^{2 Q} \delta^{jk}\partial_j\partial_k Q\right)
		+\frac{2}{3} e^{-2 Q} \delta^{i(\mu}\delta^{\nu)j} \partial_i \left(e^{-2 Q} \partial_t \partial_j Q\right) \\
		&+\frac{1}{3}e^{-2 Q} \delta^{i\mu}\delta^{\nu j}\delta_{ij} \delta^{kl} \partial_k \left(e^{-2 Q} \partial_t \partial_l Q\right)\,.
\end{split}
\end{align}
Thus, in order to compute the ensemble averaged energy momentum tensor $\mathcal{T}^{\mu\nu}$ in \eqref{E:averageT}, it remains to compute the coefficients $a$, $f_i$ and $\gamma_{ij}$ in \eqref{E:asymptotics}. In order to do so we must resort to numerics.

\section{Setting up the numerical problem}
\label{S:algorithm}
In order to numerically solve \eqref{E:EOM} and the associated initial and boundary conditions described in \eqref{E:BBFG} and \eqref{E:BCs}, we use the residual scaling symmetry $\rho \to \alpha\rho$ and $x^{\mu} \to \alpha x^{\mu}$ to set $\rho_0=1$ in \eqref{E:THawking}. Thus, all dimensionfull parameters appearing in this section are measured in units of the initial temperature. When we will present our results in the next section we will go back to dimensionfull quantities by multiplying physical quantities by appropriate powers of $\rho_0$.

The algorithm for solving \eqref{E:EOM} has been laid out in detail in \cite{Chesler:2013lia}: Given $\hat{g}_{ij}(t_0,\,\vec{x})$, $a(t_0,\vec{x})$ and $f(t_0,\vec{x})$ at some time $t_0$, we may solve the set of nested linear differential equations \eqref{E:EOM} as follows. The second order (in $\rho$) differential equation for $\Sigma$, \eqref{E:SigmaEOM} can be solved for by imposing the boundary conditions in \eqref{E:asymptotics} with $\lambda=0$. Once $\Sigma$ is known, we can solve \eqref{E:FEOM} for $F_i$ using the asymptotic behavior in \eqref{E:asymptotics}. With $F_i$ and $\Sigma$ we can solve the first order linear equation for $d_0\Sigma$, \eqref{E:d0SigmaEOM}, as an independent variable. Using \eqref{E:asymptotics} we find that $d_0\Sigma$ is uniquely determined (from the value of $a$).
We next solve \eqref{E:d0gEOM} for $d_0\hat{g}_{ij}$ with the asymptotic behavior given in \eqref{E:asymptotics} (which translates into a vanishing $\mathcal{O}(\rho)$ term in the near boundary expansion of \eqref{E:d0gEOM}). Finally, we can solve \eqref{E:AEOM} for $A$, once again with the asymptotic behavior provided by \eqref{E:asymptotics} with $\lambda=0$. 

At this point we can use the known value of $A$ to determine $\partial_t \hat{g}_{ij}$ from $d_0\hat{g}_{ij}$ and then solve the equations for $d_0F_i$ and $d_0d_0\Sigma$ to determine $\partial_t a$ and $\partial_t f_i$ which will allow us to step forward in time. In practice, once the equations for $\Sigma$, $F_i$, $d_0\hat{g}$ and $A$ are satisfied, the equations of motion for $d_0F_i$ and $d_0d_0\Sigma$ reduce to energy and momentum conservation of the boundary theory stress tensor given by \eqref{E:holographicTmn}, $\nabla_{\mu}T^{\mu\nu}=0$. Thus, we may determine the time derivatives of $a$ and $f_i$ via energy-momentum conservation.

To implement the above evolution scheme numerically, we discretize the boundary coordinates $x^i$ on a periodic grid of physical size $L$ with $N_{\parallel}$ grid points, and expand functions of $x^i$ in Fourier cardinal functions. In particular the expansion coefficients of the random force $\xi(t,\vec{x})$ are treated as stochastic functions in time, with appropriate autocorrelation functions. Spatial derivatives of $\xi$ can be computed from derivatives of the Fourier cardinal functions.

The radial coordinate, $\rho$, is taken to run in the range $0 < \rho < 2$. The metric at early times has an   apparent horizon at $\rho = 1$ (with our choice of units) and we have checked that the horizon is inside this coordinate range at later times. We have discretized the radial coordinate on a grid with $N_{\bot}$ grid points, and expanded functions of $\rho$ in  Chebyshev cardinal functions. 
Lastly, to avoid the inherently singular nature of the near boundary behavior of some of the metric components, or of the equations of motion themselves, we have found it convenient to switch from $\Sigma$, $F_i$, $d_0\Sigma$, $d_0\hat{g}_{ij}$ and $A$ to $\widetilde{\Sigma}$, $\widetilde{F}_i$, $\widetilde{d_0\Sigma}$, $\widetilde{d_0g_{ij}}$ and $\widetilde{A}$ defined via
\begin{align}
\begin{split}
\label{E:tilded}
	\Sigma =& \frac{e^Q}{\rho} + \rho^3\widetilde{\Sigma}  \\
	F_i =& \widetilde{F}_i \\
	d_0\Sigma =& \frac{e^{Q}}{2\rho^2}  + \frac{1}{2} e^{-Q} \delta^{ij}\partial_i \partial_j Q + \rho \widetilde{d_0\Sigma} \\
	d_0\hat{g}_{ij} =& \rho \widetilde{d_0 g_{ij}} \\
	A = & \frac{1}{2\rho^2} +\frac{ - \partial_t{Q} }{\rho} 
	+  \frac{1}{2} e^{-2 Q} \delta^{ij} \partial_i \partial_j Q  + \widetilde{A}\rho\,.
\end{split}
\end{align}

To summarize this section let us briefly outline our evolution scheme once more.
At each time step, $t$, we start with $\hat{g}_{ij}$, $a$ and $f_i$. We then determine the boundary value of the metric, characterized by $e^{Q(t,\vec{x})}$ by solving \eqref{E:Qtoq} and \eqref{E:dotq} and drawing $P_i(t)$ in \eqref{E:xidistribution} from a normal distribution.\footnote{In practice we have drawn $M$ values of $\vec{k}$ from the annular region at each time step, where $M$ is roughly half the number of grid points in $A$.
The same applies to \eqref{E:Qtoq}. 
}
We then solve the nested set of linear differential equations \eqref{E:SigmaEOM} to \eqref{E:AEOM} in terms of the tilde'd variables in \eqref{E:tilded}. To move forward in time we have used an Euler-Maruyama algorithm (described in appendix \ref{A:review}) using a time interval, $\Delta t$, and, as mentioned before, treated the noise using the Stratonovich scheme. At the end of the day, this algorithm provides us with $\partial_t \hat{g}_{ij}$ from which we can obtain $\hat{g}_{ij}$ at the next time step. Using energy momentum conservation we can then determine the time derivatives of $a$ and $f_i$ and use them to obtain their value at the next time step as well. We then use a $2/3$ filter to the updated values of $\hat{g}_{ij}$, $a$ and $f_i$ and repeat the procedure. With $a$, $f_i$, $\gamma_{ij}$ and $Q$ at hand, we can determine the energy momentum tensor at all times via \eqref{E:holographicTmn}. An approximation to the ensemble averaged stress tensor $\mathcal{T}^{\mu\nu}$ is obtained by repeating this algorithm a large, $N_0$, number of times and averaging over the resulting stress tensors obtained in each instance.

\section{Results}

The expectation value of the stress tensor obtained via the algorithm outlined in \ref{S:algorithm} may depend, a priori, on the parameters of the noise, $\tau$, $k_f$, $\Delta k$ and $D$ in  \eqref{E:xidistribution} as well as on the length of the spatial torus $L$. In the current text we have focused on
\begin{equation}
\label{E:ourparameters}
	\frac{\tau}{\rho_0} = \frac{1}{4}\,,
	\quad
	\frac{L k_f}{2\pi} = 21 \,,
	\quad
	\frac{L \Delta k}{2\pi} = \frac{3}{4}\,,
	\quad
	\frac{D}{\rho_0} = \frac{3}{1250000}\,,
	\quad
	\frac{L}{\rho_0} = 500\,.
\end{equation}
We have chosen \eqref{E:ourparameters} since, after some exploration, these, as well as their nearby values, allow for a stress tensor whose associated energy power spectrum has the Kolmogorov scaling law that we're interested in.  A full study of the dependence of the expectation value of the stress tensor on these parameters (and other more general noise terms) is left for future work.  

In order to implement the aforementioned algorithm numerically we must also choose an appropriate spatial and temporal grid. We have found that the somewhat modest
\begin{equation}
\label{E:gridparameters}
	N_{\bot} = 70, \quad N_{\parallel} = 11, \quad \frac{\Delta t}{\rho_0} = \frac{1}{40}
\end{equation}
where sufficient to demonstrate Kolmogorov scaling with an inertial range which spanned a decade of momenta. We plan on exploring the exact dependence of turbulent features of the energy momentum tensor on the parameters in \eqref{E:ourparameters} in future work as well. In the remainder of this section we present our results for the stress tensor and the dual geometry associated with it given the parameters in \eqref{E:ourparameters} and \eqref{E:gridparameters}.

\subsection{The boundary stress tensor}

\begin{figure}[hbt]
\centering
	\includegraphics[width=0.8\textwidth]{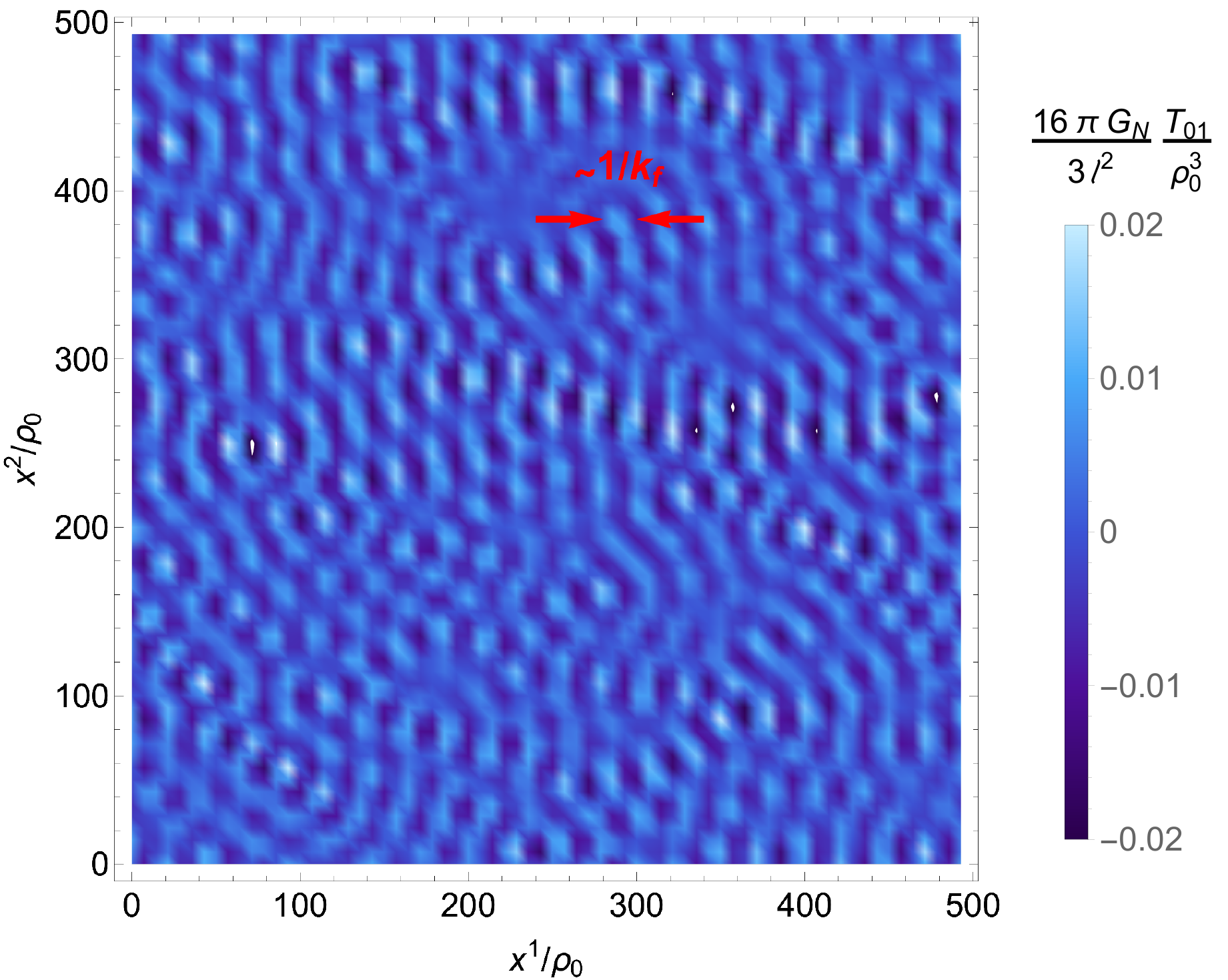}
	\caption{\label{F:typicalstress} A snapshot of the energy flux, $T_{01}(t,x^1,x^2)$ at $t=750 \rho_0$ where $\rho_0 = 4\pi T_0/3$ and $T_0$ is the temperature prior to the onset of the driving force. The typical scale we observe is of the same order as the driving scale, $1/k_f$.}
\end{figure}

\begin{figure}[hbt]
\centering
	\begin{minipage}{0.45\textwidth}
	\centering
	\includegraphics[width=\textwidth]{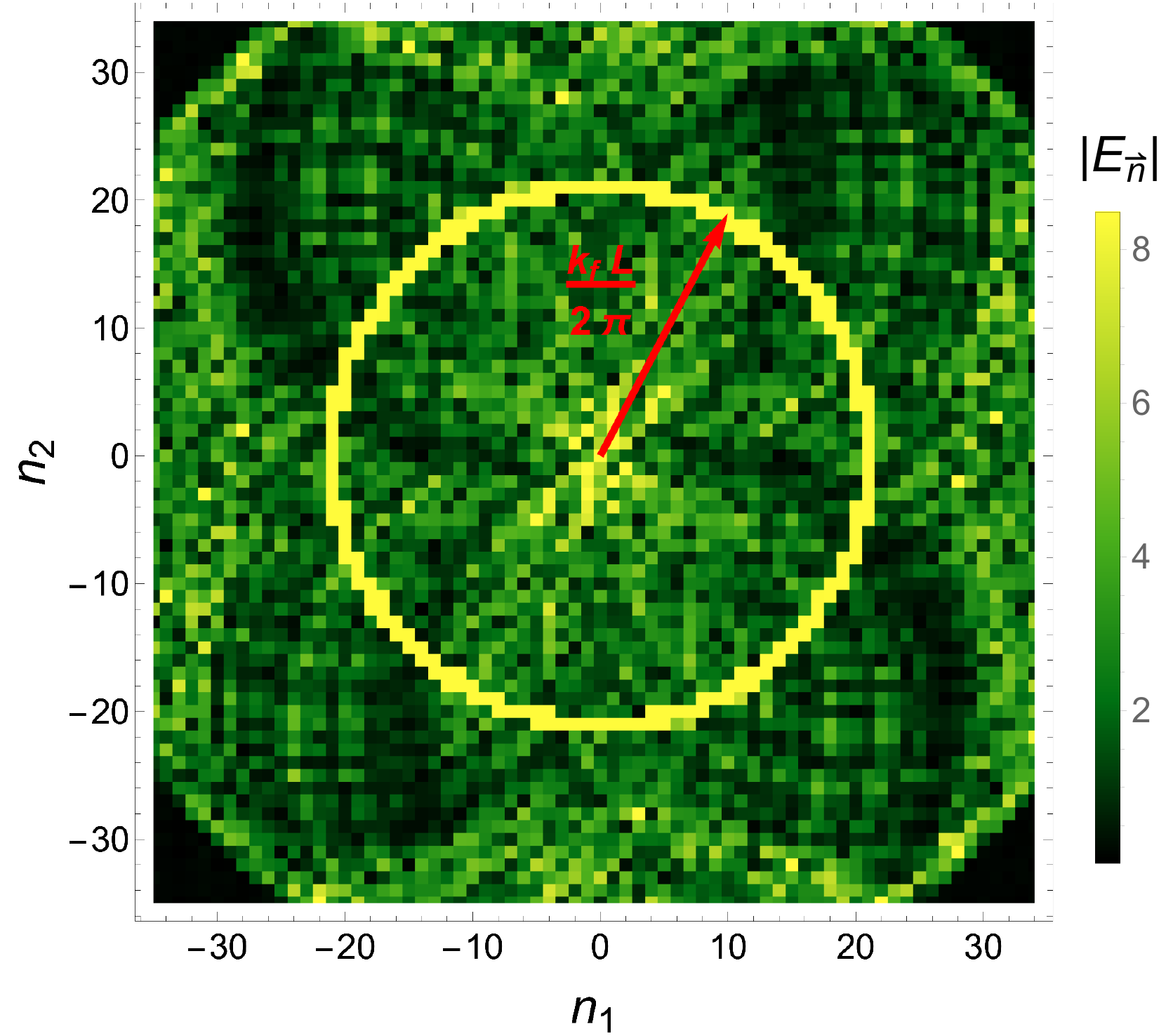}
	\end{minipage}
	\hfill
	\begin{minipage}{0.5\textwidth}
	\includegraphics[width=\textwidth]{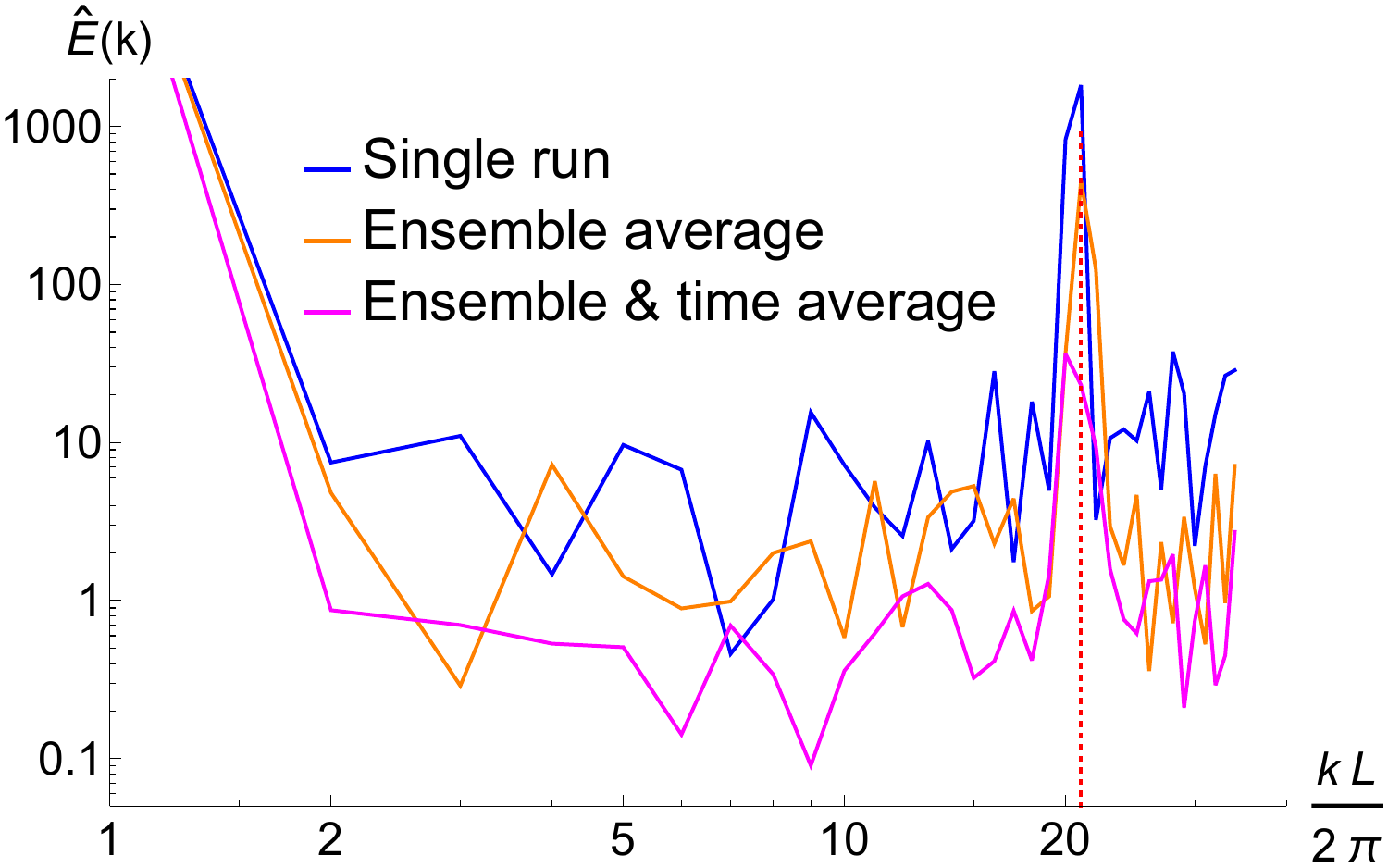}
	\end{minipage}
\\
	\begin{minipage}{0.45\textwidth}
	\centering
	\includegraphics[width=\textwidth]{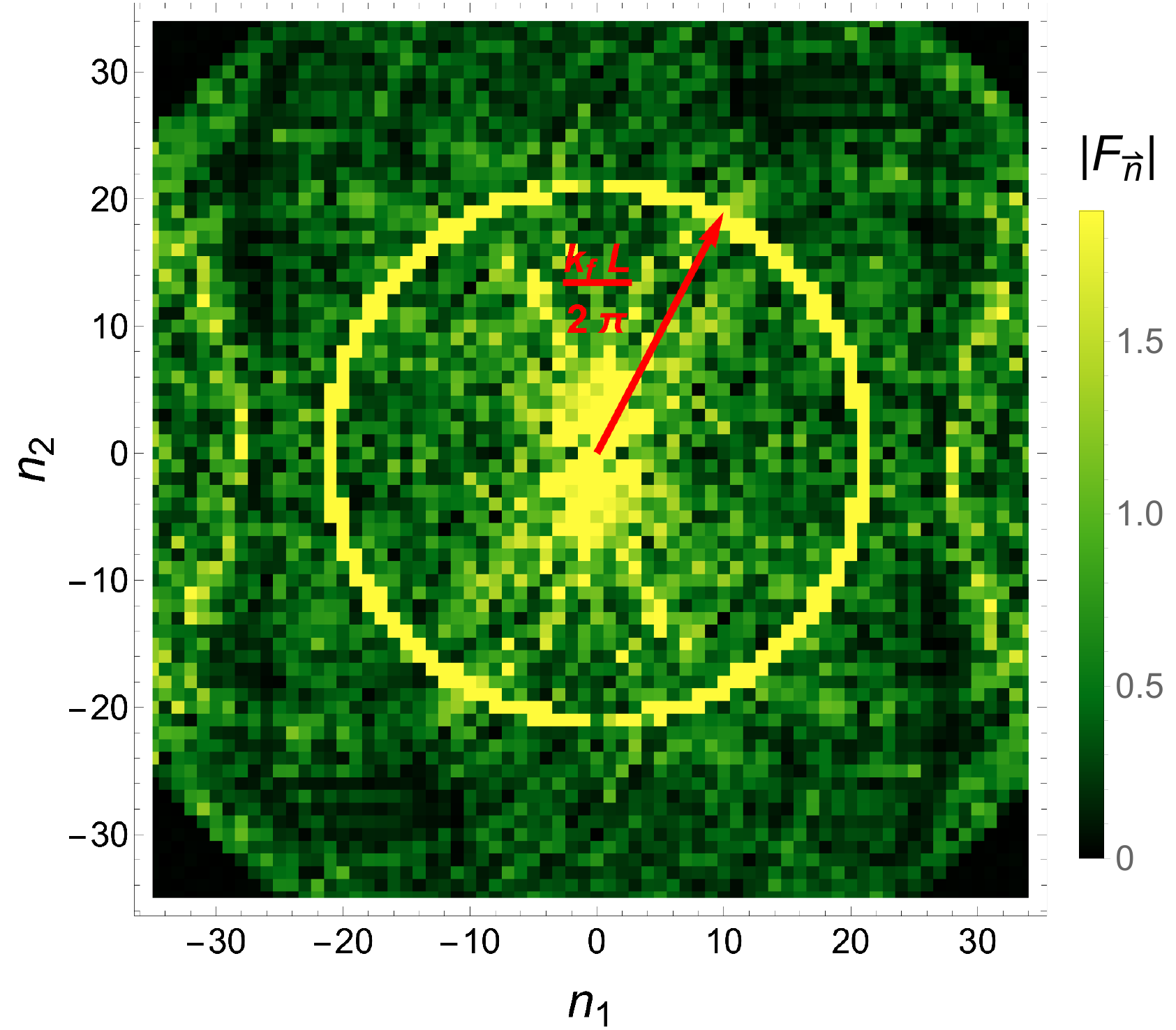}
	\end{minipage}
	\hfill
	\begin{minipage}{0.5\textwidth}
	\includegraphics[width=\textwidth]{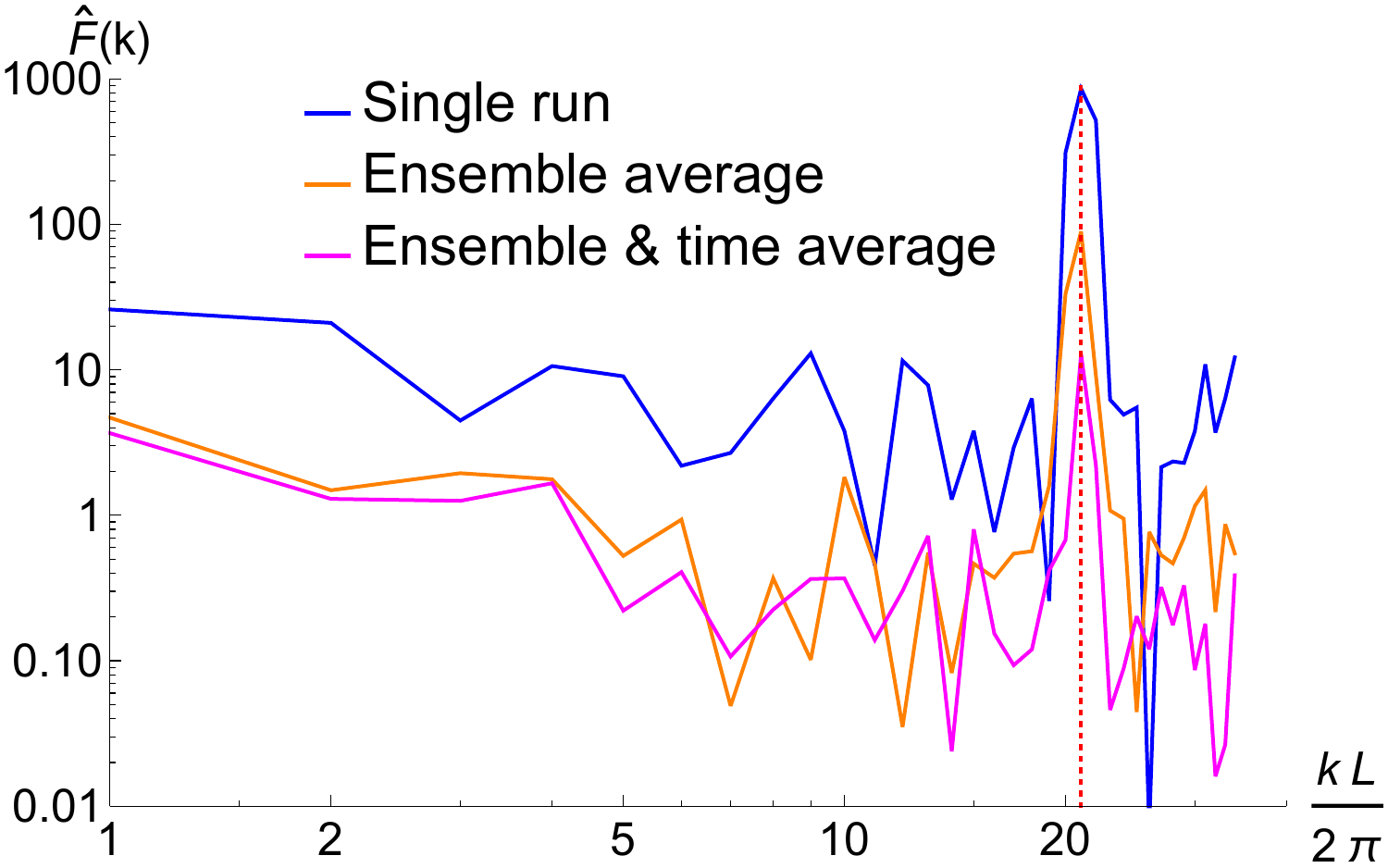}
	\end{minipage}
	\caption{\label{F:typicalstressfourier} A set of plots exhibiting the behavior of the Fourier decomposition of the energy density (top) and energy flux (bottom), $\frac{16 \pi G_N}{3\ell^2} {T}_{00}(t,\,\vec{x}) = \frac{1}{L^2\rho_0} \sum_{\vec{n}} E_{\vec{n}}e^{i\frac{2 \pi \vec{n}}{L} \cdot \vec{x}} $ and $\frac{16 \pi G_N}{3\ell^2} {T}_{01}(t,\,\vec{x}) = \frac{1}{L^2\rho_0} \sum_{\vec{n}} F_{\vec{n}}e^{i\frac{2 \pi \vec{n}}{L} \cdot \vec{x}} $,
at $t=1125 \rho_0$ with $\rho_0 = 4\pi T_0/3$ and $T_0$ the initial temperature. Plots of $|E_{\vec{n}}|$ and $|F_{\vec{n}}|$ appear on the left column. The centered ring is located at the driving scale, $k_f$.  In  the right column we have plotted the (dimensionless) angle averaged energy density $\hat{E}(|\vec{k}|) = \int E_{\frac{\vec{k}L}{2\pi}} \frac{|\vec{k}| L}{2\pi}  d\theta$ and flux $\hat{F}(|\vec{k}|) =\int F_{\frac{\vec{k}L}{2\pi}} \frac{|\vec{k}| L}{2\pi}  d\theta$ (where $E_{\vec{n}}$ and $F_{\vec{n}}$ have been analytically continued to all values of $\vec{n}$) for a single run (blue), an average over 36 independent runs (orange) and a further average over 19 nearby time steps (pink). The dashed red line corresponds to the driving scale $k_f$.}
\end{figure}

Once the stochastic metric is turned on, the components of the energy momentum tensor $\hbox{Tr}\left(\varrho T^{\mu\nu} \right)$ quickly become dominated by the driving force. A snapshot exhibiting typical behavior of the stress tensor can be found in figure \ref{F:typicalstress}.
After we ensemble averaged the expectation value of the stress tensor over 36 independent runs, we found that the magnitude of the oscillations resulting from the driving force decreased by an order of magnitude. Carrying out a time average over 19 snapshots, the oscillations decreased further. To demonstrate this effect more clearly we have plotted the Fourier components of the stress tensor and their angle averaged value in figure \ref{F:typicalstressfourier}. 

Apart from the clear signature of the driving force observed in figure \ref{F:typicalstressfourier}, we also note a possible underlying structure associated with momentum modes lower than the driving scale. Momentum modes larger than the driving scale seem to contain modes larger than those available on the torus and are therefore unreliable. In future work we will treat these modes properly.

Turbulent flow in non relativistic fluids is characterized by power law behavior of correlation functions of the velocity field, $\vec{v}$. In particular, a hallmark of incompressible turbulent flow is that the velocity field correlation function $\overline{v^i(\vec{x}+\vec{r})v^j(\vec{x})}$ satisfies
\begin{subequations}
\label{E:K41realspace}
\begin{equation}
	\overline{v^i(\vec{x}+\vec{r})v^j(\vec{x})} = G(r) r^i r^j  + H(r) \delta^{ij}
\end{equation}
with
\begin{equation}
	G(r) \sim r^{-{\frac{4}{3}}}
	\qquad
	F(r) \sim r^{\frac{2}{3}}
\end{equation}
\end{subequations}
in a certain inertial range of scales. This behavior implies that the energy power spectrum defined via
\begin{equation}
\label{E:PSdef}
	\hat{\epsilon}(t,\,k) = \frac{1}{(2\pi)^2} \int \frac{1}{2} \delta_{ij}  \overline{\hat{v}^i \hat{v}^j} k  d\theta
\end{equation}
with
\begin{equation}
	\hat{v}(t,\,\vec{k}) = \int \vec{v}(t,\,\vec{x}) e^{-i \vec{k} \cdot \vec{x}} d^2x
\end{equation}
satisfies the scaling law
\begin{equation}
\label{E:Kolmogorov}
	\hat{\epsilon} \sim k^{-\frac{5}{3}}\,,
\end{equation}
in the inertial range.
To be slightly more precise, the power law behavior of $\hat{\epsilon}$ behaves as in \eqref{E:Kolmogorov} for momentum scales smaller than the forcing scale, $k_f$, and as $k^{-3}$ for momenta larger than the forcing scale (assuming that the forcing scale is not too high). The former scaling law is associated with what is usually referred to as an inverse energy cascade and the latter to a direct enstrophy cascade. Such behavior of the power spectrum for the velocity field is unique to $2+1$ dimensional flow and is related to the existence of an accidentally conserved enstrophy charge. In $3+1$ dimensional flow enstrophy is not conserved and one observers a power law behavior of the form \eqref{E:Kolmogorov} only at momenta above the forcing scale---a direct energy cascade.

The scaling law described in \eqref{E:Kolmogorov} is, strictly speaking, valid for non relativistic fluid flow on a manifold with a Euclidean geometry. There are a handful of works which attempt to generalize \eqref{E:Kolmogorov} to non relativistic systems in such a way that the non relativistic limit of this generalization will reduce to \eqref{E:Kolmogorov} \cite{Fouxon:2009rd,Carrasco:2012nf,Westernacher-Schneider:2015gfa,Westernacher-Schneider:2017snn,Saridthesis}. In particular, the authors of \cite{Westernacher-Schneider:2017snn,Saridthesis} have numerically verified the predictions of \cite{Fouxon:2009rd,Westernacher-Schneider:2015gfa} for relativistic fluid flow at low Mach numbers. Unfortunately, for the theories considered, low Mach number implies velocities of at most $1/\sqrt{2}c$ at which relativistic corrections are still small. To obtain truly relativistic turbulent flow, one would need to consider supersonic flow which is notoriously challenging to simulate. It may well be that relativistic turbulent flow is not universal \cite{Eyink:2017zfz}. A study of the relativistic entropy current can be found in \cite{Carrasco:2012nf,Marjieh:2020odp,Pinzani-Fokeeva:2021klb}.

\begin{figure}[hbt]
\centering
	\includegraphics[width=0.8\textwidth]{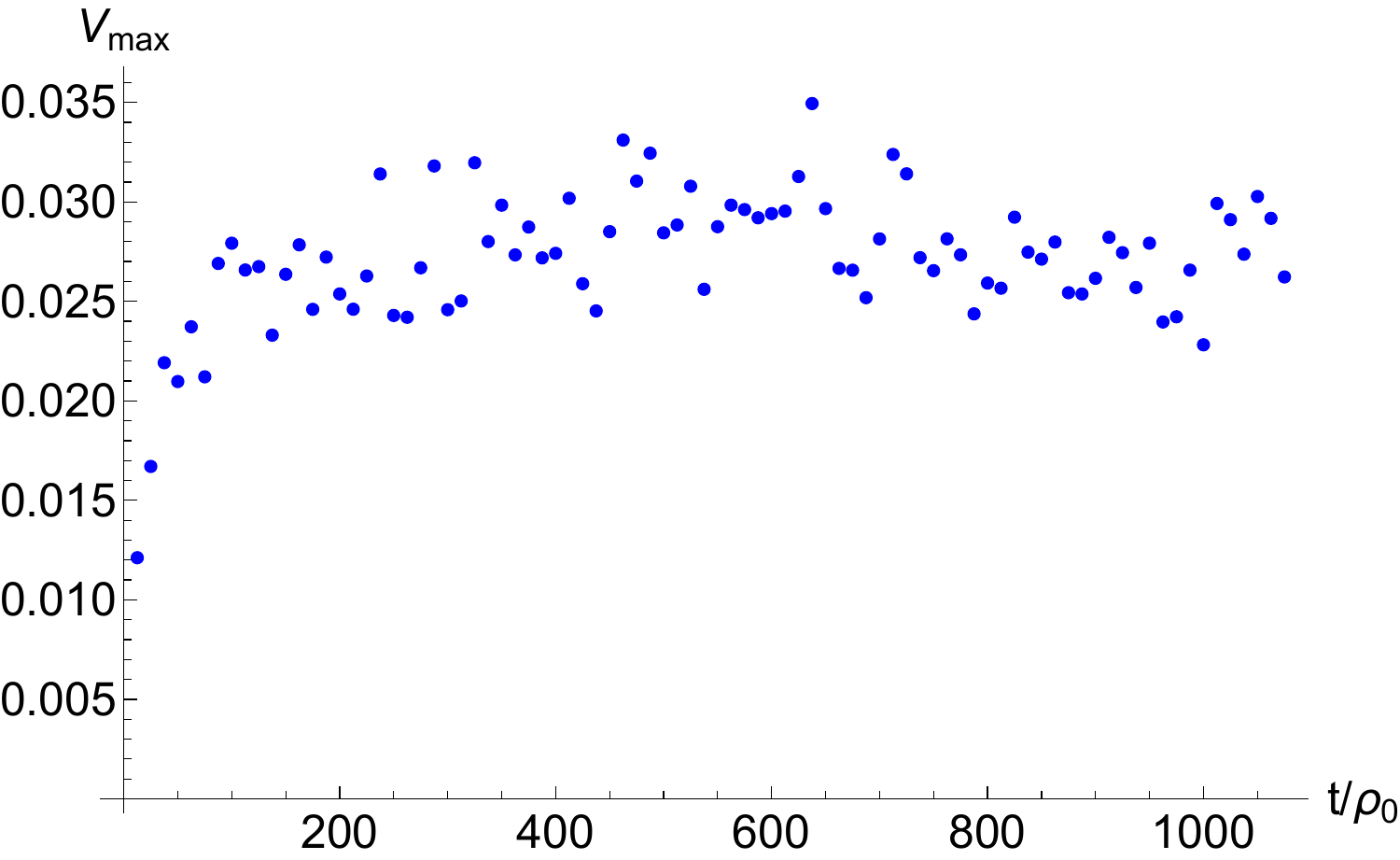}
	\caption{\label{F:velocities} Maximal values of the velocity field (relative to the speed of light) as a function of time for a single run (no ensemble averaging). }
\end{figure}

Be that as it may, in order to compare our result for the expectation value of the stress tensor to turbulent behavior, we identify the (Landau frame) velocity field $u_L^{\mu}$ and proper energy density $\epsilon_L$ of the fluid, via the eigenvalue equation
\begin{equation}
	\hbox{Tr}\left(\varrho T^{\mu}{}_{\nu}\right) u_L^{\nu} = -\epsilon_L u_L^{\nu}\,,
\end{equation}
where the velocity is normalized such that $u_L^{\mu}u_{L\,\mu}=-1$. We define the velocity field $\vec{v}$ via $v^i = u^i/u^0$. As can be observed in figure \ref{F:velocities} the velocities attained in a typical run are far from being relativistic.
Since the relativistic formalism developed in \cite{Fouxon:2009rd,Carrasco:2012nf,Eyink:2017zfz} reproduces the non relativistic result \eqref{E:Kolmogorov} in the non relativistic limit, we will content ourselves with studying the non relativistic power spectrum and its relation to the scaling behavior given in \eqref{E:Kolmogorov}. 

\begin{figure}[hbt]
\centering
	\includegraphics[width=0.8\textwidth]{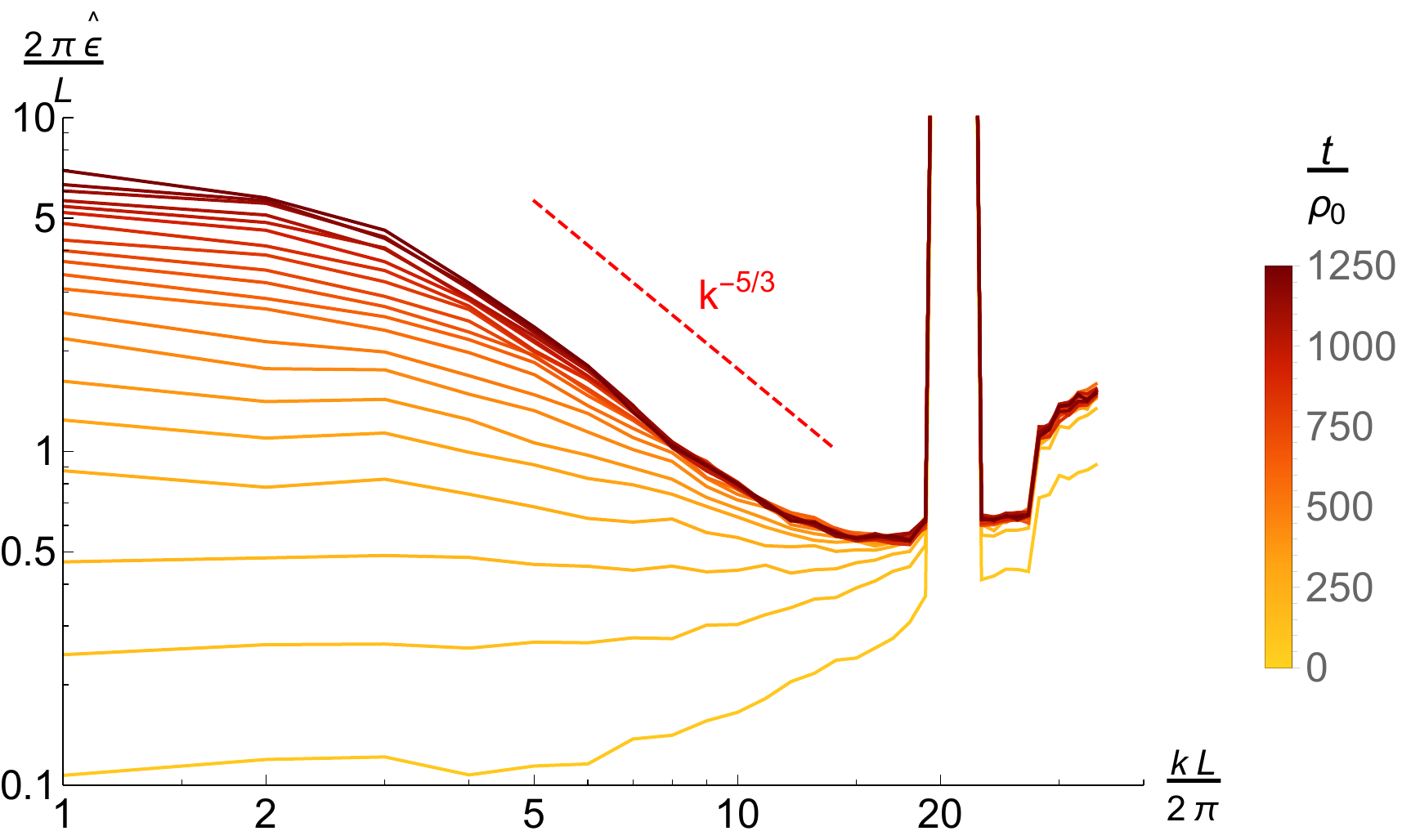}
	\caption{\label{F:powerspectrum} The non relativistic energy power spectrum defined in \eqref{E:PSdef} as a function of momenta for different times. The displayed data has been obtained after ensemble averaging over 36 runs. The clearly visible maximum is due to the driving force located at $k_f L / 2\pi = 21$, and has been clipped in order to highlight the power law behavior at low momenta.}
\end{figure}

At this point the astute reader might correctly point out that the non relativistic energy power spectrum for the case at hand is not quite \eqref{E:PSdef} due to the stochastic component of the spatial metric. Still, the quantity $\hat{\epsilon}$ defined in \eqref{E:PSdef} will satisfy \eqref{E:Kolmogorov} due to \eqref{E:K41realspace}. In what follows, for lack of a better name, we will continue to refer to $\hat{\epsilon}$ as the non relativistic energy power spectrum. A straightforward computation of $\hat{\epsilon}$ yields a power law of the form $k^{-\nu}$ with $\nu = 1.53 \pm 0.05$ in the inertial range $5 < \frac{k L}{2\pi} < 14$.  
See figure \ref{F:powerspectrum}.

\subsection{The bulk metric}
\begin{figure}[hbt]
\centering
	\includegraphics[width=0.8\textwidth]{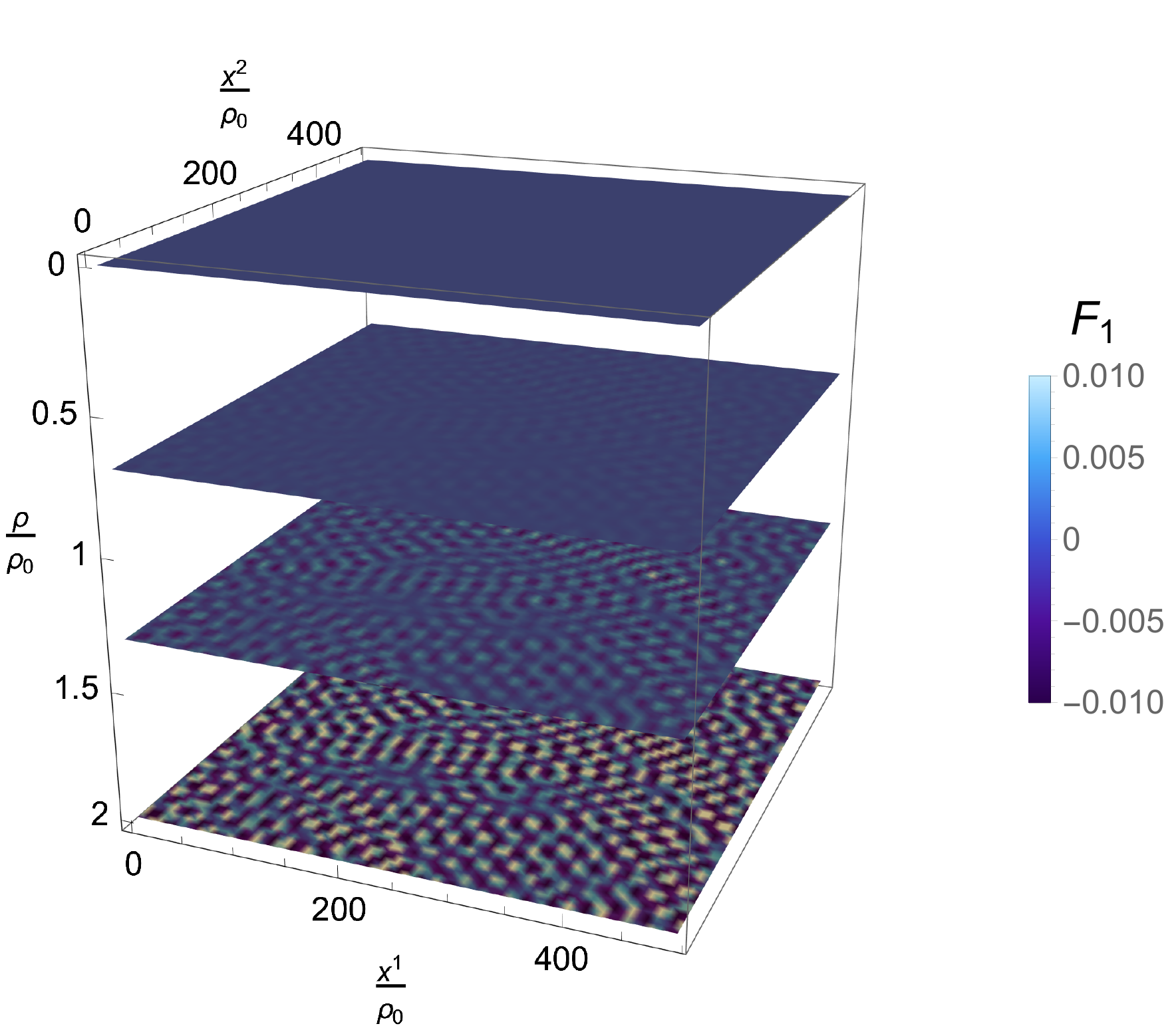}
	\caption{\label{F:BulkF1} A stacked density plot of the $t$, $x^1$ component of the metric, $F_1(\rho,t,\vec{x})$, as a function of the radial coordinate $\rho$ at time $t=1125 \rho_0$, ensemble averaged over 36 runs. As the radial coordinate $\rho$ increases (moves farther from the boundary) the magnitude of the perturbations induced by the driving force become larger. In each run there exists an apparent horizon at $\rho=\rho_h(t,\vec{x})$ with $0.9 \lesssim \rho_h \lesssim 1.1 $.}
\end{figure}

\begin{figure}[hbt]
\centering
	\includegraphics[width=0.7\textwidth]{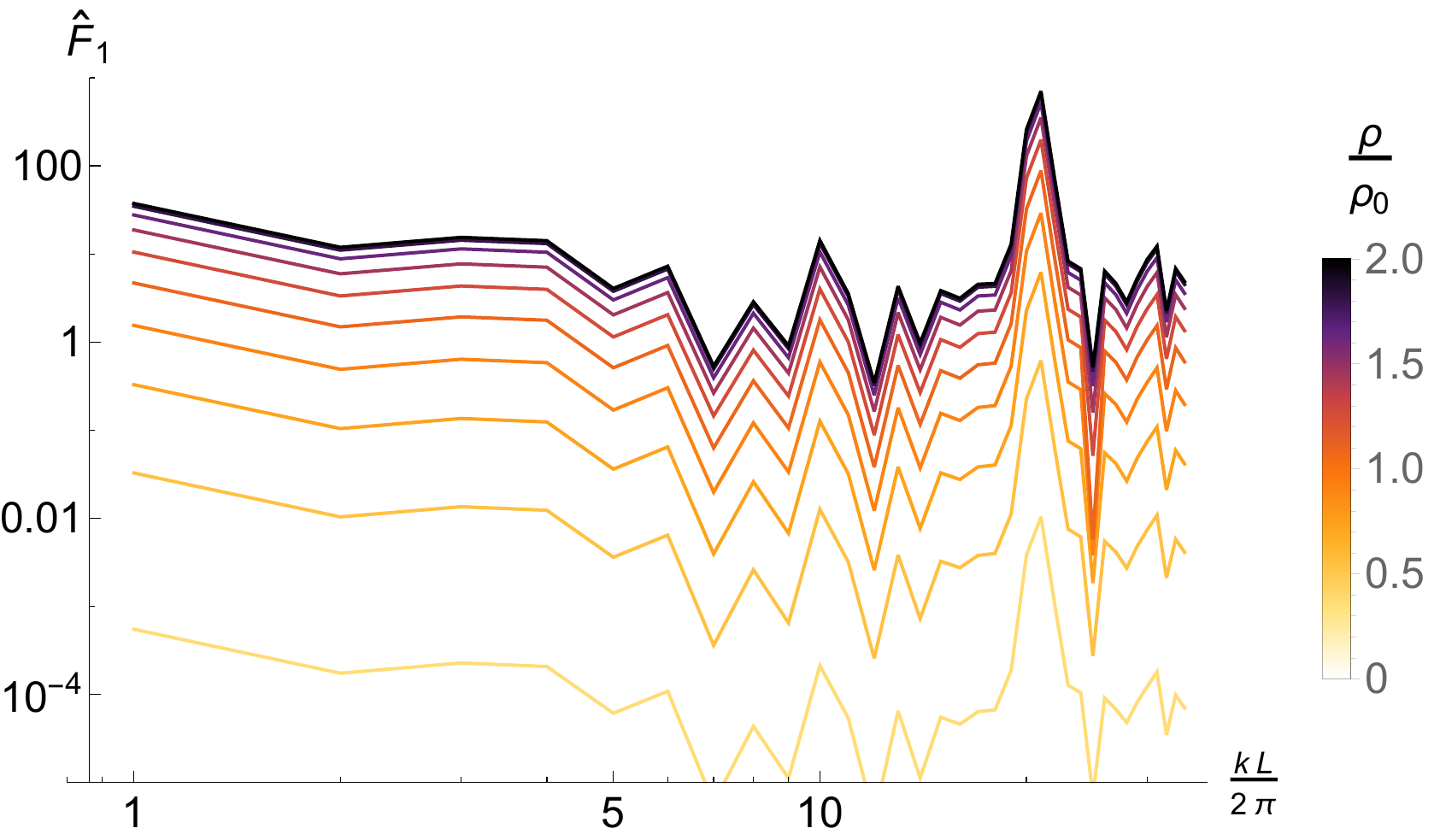}
	\caption{\label{F:FourierG} A plot of the Fourier decomposition of $F_1(\rho,t,\vec{x}) = \frac{1}{L^2}\sum_{\vec{n}} F_{1\,\vec{n}}(\rho,t) e^{i \frac{2\pi \vec{n}}{L}\cdot \vec{x}}$, the $t$, $x^1$, component of the metric, averaged over angular directions, $\hat{F}_1(\rho,t) =\frac{1}{\rho_0^2} \int F_{1\,\frac{|\vec{k}|L}{2\pi}}(\rho,t) \frac{|\vec{k}|L}{2\pi}  d\theta$, at time $t=1125 \rho_0$ and ensemble averaged over 36 runs. As the radial coordinate $\rho$ increases (moves farther from the boundary) the magnitude of the perturbations induced by the driving force become larger.}
\end{figure}

As discussed in the introduction, one of the main goals of this program is to understand how the turbulent behavior exhibited in figure \ref{F:powerspectrum} manifests itself in terms of geometric quantities. As a general rule, perturbations on the boundary of AdS are expected to inflate as one moves towards its interior due to the AdS warp factor. In figure \ref{F:BulkF1} we have plotted a snapshot of the behavior of $F_1$ (the $t$, $x^1$ component of the bulk metric) as a function of the spatial coordinates. We find that the stochastic behavior of the boundary metric dominates the behavior of $F_1$ and its overall amplitude increases as a function of the radial coordinate $\rho$.
A more quantitative demonstration of the same feature can be found in figure \ref{F:FourierG} where we have plotted the Fourier components of $F_1$ as a function of $\rho$.

The area element of the apparent horizon follows a similar trend. Following the construction of \cite{Chesler:2013lia}, the location of the apparent horizon, $\rho=\rho_h(t,\vec{x})$ may be determined by requiring that $\rho - \rho_h = 0$ is a null surface with zero expansion. On a typical run $\rho_h$ will have features similar to those exhibited in the density plot in figure \ref{F:typicalstress} with $\rho_h$ within less than 10\% of unity.

Once $\rho_h$ is known, the horizon area element can be read off of
\begin{equation}
	\mathcal{A}(t,\vec{x}) = \sqrt{g_h(\rho_h(t,\vec{x}),t,\vec{x})} = \Sigma(\rho_h(t,\vec{x}),t,\vec{x})^2\,.
\end{equation}
As it turns out, $\Sigma$ is given to a good approximation by its boundary value given in \eqref{E:asymptotics}, 
\begin{equation}
	\Sigma \sim \frac{e^{Q}}{\rho}\,,
\end{equation}
and the resulting Fourier decomposition of the horizon area element,
\begin{equation}
	\mathcal{A} = \frac{1}{L^2} \sum_{\vec{n}} \mathcal{A}_{\vec{n}} e^{i \frac{2\pi \vec{n}}{L}\cdot \vec{x}}\,,
\end{equation}
takes a form similar to that exhibited in on the left column of figure \ref{F:typicalstressfourier}.

\begin{figure}[hbt]
\centering
	\includegraphics[width=0.7\textwidth]{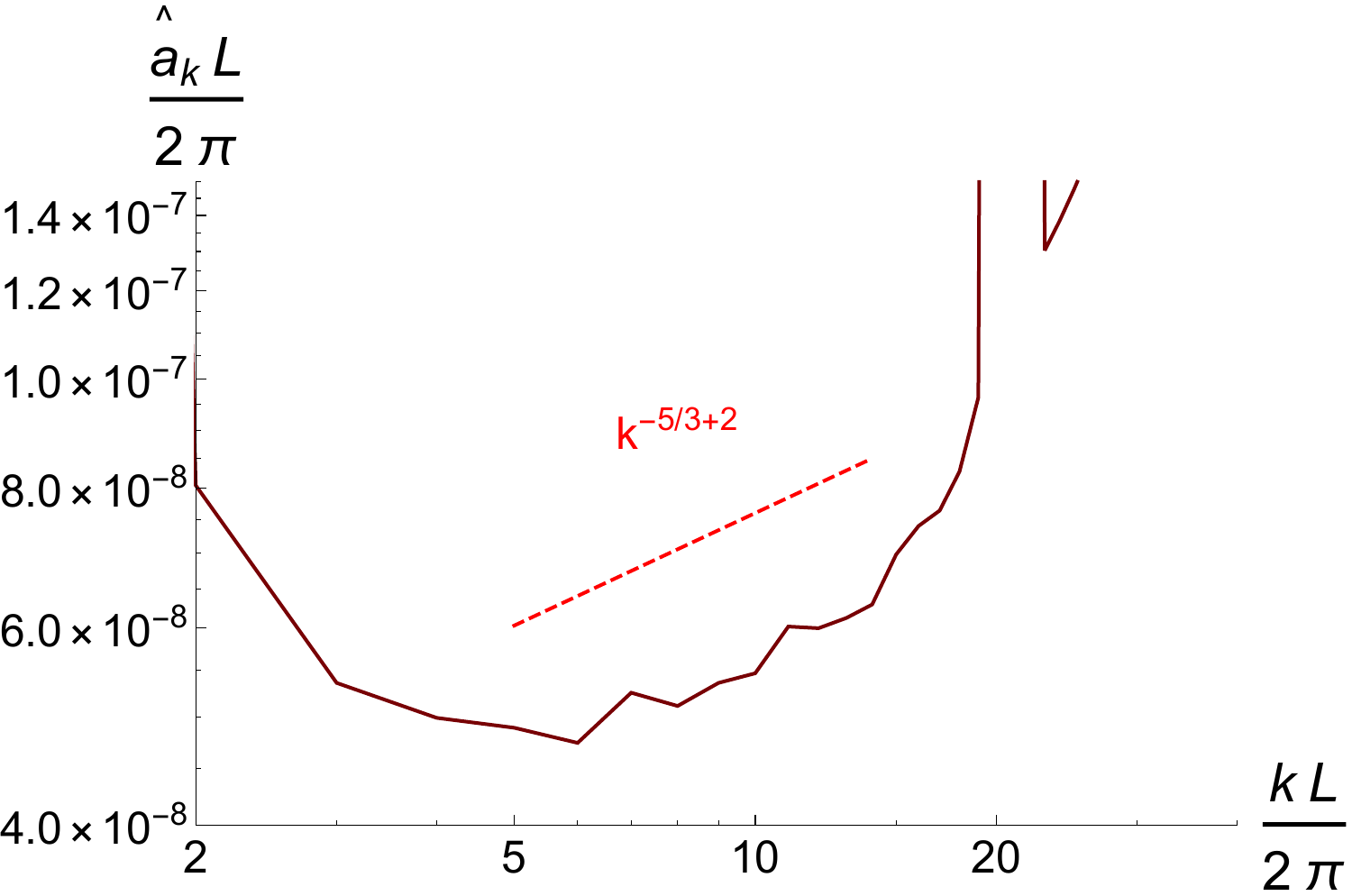}
	\caption{\label{F:hps} A plot of the horizon area power spectrum, defined in \eqref{E:HPS}, at time $t=1250 \rho_0$ as a function of momentum, ensemble averaged over 36 runs. The fit to a $-5/3+2$ power law is inline with other indications for this scaling described in the literature and discussed in the main text. }
\end{figure}

Instead of considering the Fourier components of $\mathcal{A}$, one may consider a horizon area power spectrum defined by
\begin{equation}
\label{E:HPS}
	\hat{a}_{k} = \frac{1}{2\pi L} \sum_{|\vec{n}| = \frac{k L}{2\pi}} |\Sigma_{\vec{n}} |^2 \rightarrow 
	\frac{1}{(2\pi)^2}  \int \Big|\Sigma_{\frac{\vec{k}L}{2\pi}}\Big|^2 k  d\theta\,,
\end{equation}
where 
\begin{equation}
	\Sigma = \frac{1}{L^2} \sum_{\vec{n}} \Sigma_{\vec{n}} e^{i \frac{2\pi \vec{n}}{L}\cdot \vec{x}}
\end{equation}
and the arrow denotes the continuum limit.
The horizon area power spectrum is related to the horizon area via
\begin{equation}
	A = \int \mathcal{A}d^2x = \frac{2\pi}{L} \sum_n \hat{a}_{\frac{2\pi n}{L} }\rightarrow \int \hat{a}_k dk\,,
\end{equation}
similar to the relation between the velocity field and the non relativistic energy power spectrum, c.f., \eqref{E:PSdef}. In the context of decaying turbulence, the authors of \cite{Adams:2013vsa} observed that the horizon power spectrum follows the non relativistic power spectrum to good accuracy. In \cite{Rozali:2017bll} it was shown that (in the limit of a large number of dimensions) $\hat{a}$ is proportional to $\hat{\epsilon}k^{2}$ but only for low Mach number. Here, we make a similar observation. The apparent horizon power spectrum closely follows $\hat{\epsilon}$ and obeys a scaling relation $\hat{a} \sim k^{-\nu}$ with $\nu = -0.28 \pm 0.03$, see figure \ref{F:hps}.

\section{Discussion}
\label{S:discussion}

In this work we have demonstrated that a holographic dual to a field theory place on an external random metric is well defined provided that the driving force is sufficiently differentiable in a sense described in detail in section \ref{S:stochastic}. We then constructed a bulk dual to an initially thermalized state driven out of equilibrium once the background metric fluctuated randomally. The resulting energy momentum tensor exhibited a power spectrum which agrees nicely with turbulent behavior, as exhibited in figure \ref{F:powerspectrum}.

In figure \ref{F:powerspectrum} as well as figures \ref{F:typicalstress} and \ref{F:typicalstressfourier} which exemplify the behavior of the stress tensor, we observe that the dominant excitation of the fluid is the driving force---it is several orders of magnitude larger than any other noteable feature of the fluid. We expect that if one waits long enough then energy will transfer from the driving scale to the lower momentum modes which will dominate the power spectrum. Unfortunately, in the absence of large scale friction, the energy density will eventually pile up in low momentum modes spoiling the Kolmogorov scaling behavior observed in figure \ref{F:powerspectrum}. A larger inertial range, which neccessitates larger grid sizes, may ameliorate this behavior.

In numerical simulations of $2+1$ dimensional, incompressible, flow, a turbulent steady state is generated by adding large scale friction which removes excess energy from low momentum modes. More precisely, energy is pumped into the system at the driving scale, $1/k_f$, and gets dissipated at large scales (by, say, friction with the boundary). Once a steady state is reached a condensate will form. While it would be appealing to generate such a condensate in a setup of the type discussed in this work, it is difficult to envision how large scale friction may be incorporated in a natural way within the framework of holography. One possibility is to induce appropriate boundary conditions on the boundary manifold possibly following the line of reasoning in \cite{Fujita:2011fp}.

From the bulk point of view, the boundary excitations of the metric propagate into the interior with an amplified amplitude due to the AdS warp factor. See figure \ref{F:BulkF1}. Thus, if a signature of Kolmogorov scaling is apparent in the components of the energy momentum tensor, it will also be apparent on the {event} horizon. It is difficult to judge from the plots in figure \ref{F:typicalstressfourier} whether the ensemble averaged energy density $\mathcal{T}_{00}$ exhibits an underlying scaling behaviour. If we use the leading order hydrodynamic constitutive relations to evaluate the energy density, and expand it around a low velocity solution, we find
\begin{equation}
	\mathcal{T}_{00} = \overline{ \epsilon u_{0}u_{0}} =\overline{ \epsilon} + \overline{ \epsilon v^2} + \ldots
\end{equation}
reminiscent of the non relativistic energy density $\overline{v^2}$. Further, in a conformal theory (and in the absence of additionally conserved charges) the energy density $\epsilon$ is related to the entropy density $s$ via $s \propto \epsilon^{\frac{d-1}{d}}$ and the latter should be related to an area power spectrum of the event horizon, similar to the one defined in \eqref{E:HPS}. Following \cite{Bhattacharyya:2008xc} (see also \cite{Adams:2013vsa}) the event horizon and apparent horizon almost coincide as long as hydrodynamics is a good approximation to the behavior of the energy momentum tensor. While the power law behavior of the horizon area power spectrum is  in line with the findings of \cite{Adams:2013vsa,Rozali:2017bll} for non relativistic theories, it is unclear how or whether such a relation holds for  relativistic flow, and whether $\overline{\epsilon}$ captures any of its features. Unfortunately, our current grid is too small to resolve these issues. 

Increasing the grid size and the inertial range in an attempt to resolve the behavior of the black brane in the vicinity of its horizon, is vital to understanding the role of turbulence in black hole dynamics. However, before doing so, we would like to step back and understand the Reynolds number, or its analog, in the type of flows we are considering. In non relativistic, incompressible, fluid flow the Reynolds number $Re$ is given by $Re=v X / \nu$ where $v$ is the typical scale for the velocity of the fluid, $X$ is a typical length scale and $\nu$ is the kinematic viscosity. When the Reynolds number is much larger than one, typically of order of thousands, the flow becomes turbulent. The Reynolds number, or its equivalent in  relativistic fluid flow is less studied. In \cite{Fouxon:2009rd} it was suggested that in a relativistic setting $Re = \frac{T\,X}{\eta/s}$ which is the ratio of the ideal to viscous energy momentum tensor. Here $T$ is a typical temperature scale, $X$ a typical length scale and $\eta/s$ the shear viscosity to entropy density ratio. Given that $\eta/s$ is fixed in a holographic setting, we should be able to increase the size of the inertial range by modifying $X$ or $T$. True universality would imply that changing parameters associated with the driving force, such as its amplitude $D$, its length scale $1/k_f$ or its decay time $\tau$ should not affect the flow. It would be satisfying if this feature or its breaking can be demonstrated explicitly. 

Even if the grid size is resolved and the ensemble averaged event horizon is found to have, in some sense, a fractal dimension of the type advocated in \cite{Adams:2013vsa} or a variant of it, c.f., \cite{Westernacher-Schneider:2017xie}, its geometric implication is still somewhat murky. It is unclear whether ensemble averaging over classical black hole geometries gives a meaningful geometric structure, or whether at all one should ensemble average over the event horizon or, instead, consider the event horizon of the ensemble averaged geometries. On a related note, perhaps it is possible to generate stochastic noise by other means using, say, the constructions of \cite{Glorioso:2018mmw,deBoer:2018qqm} which generate the Schwinger Keldysh generating function, naturally equipped with a stochastic term. We plan on dealing with these, and other related issues in the near future.

\section*{Acknowledgements}
We thank A. Frishman and Y. Kafri for useful discussions. AY and SW are supported in part by an Israeli Science Foundation excellence center grant 2289/18 and a Binational Science Foundation grant 2016324.

\begin{appendix}

\section{Review of SDE's}
\label{A:review}
There is an extensive body of  literature on stochastic differential equations to which this extremely brief review can not do justice. In what follows we focus on those features of SDE's most relevant to the current work. We refer the reader to the many textbooks on the subject, e.g., \cite{Oskendal,gardiner2009stochastic}, for a more detailed exposition.

Consider the ordinary stochastic differential equation,
\begin{equation}
\label{E:SDEgeneral}
	\frac{\partial}{\partial t} X(t) = g(X(t)) + h(X(t))\xi(t)\,,
\end{equation}
where $\xi(t)$ is a random variable. As it stands, the random nature of $\xi(t)$, implies that \eqref{E:SDEgeneral} is ill defined. To see this, let us rewrite \eqref{E:SDEgeneral} in integral form
\begin{equation}
\label{E:SDEintegral}
	X(t) = X(t_0) + \int_{t_0}^t g(X(t')) dt' + \int_{t_0}^t h(X(t')) \xi(t')dt'\,.
\end{equation}
If we identify the integrals on the right hand side of \eqref{E:SDEintegral} as Riemann-Stieltjes integrals, then the second term on the right reads
\begin{align}
\begin{split}
\label{E:defSI}
	\int_{t_0}^t h(X(t'))\xi(t')dt' = \lim_{\Delta t  \to 0} \sum_{n=0}^{N-1} h(X(\tau_n)) \int_{t_0+n \Delta t}^{t_0+(n+1)\Delta t} \xi(t')dt'
\end{split}
\end{align}
with $N \Delta t = t-t_0$ and $t_0+n\Delta t \leq \tau_n < t_0+(n+1)\Delta t$. 
Intorducing $t_n = t_0 + n \Delta t$ and denoting $\tau_n = \cos^2 \alpha_n t_{n+1} +\sin^2 \alpha_n t_n$ we see that the average, $\overline{ \int_{t_0}^t h(X(t')) \xi(t') dt' }$, will depend on $\alpha_n$ via the difference between $\overline{X(t_n)\xi(t')} $ and $\overline{X(t_{n+1})\xi(t') }$ with $t_n < t' < t_{n+1} $. Thus, equation \eqref{E:SDEgeneral} is ill defined, and as opposed to ordinary (non stochastic) differential equations it must be supplemented with a prescription for carrying out the sum in \eqref{E:defSI}. 

One often used prescription for defining the integral in \eqref{E:defSI} is referred to as an It\^{o} prescription where
\begin{equation}
\label{E:defIto}
	\tau_n = t_n
\end{equation}
(or $\alpha_n = \pi/2$). Alternately, the Stratonovich prescription is given by
\begin{align}
\begin{split}
\label{E:defStra}
	\int_{t_0}^t h(X(t'))\xi(t')dt' = \lim_{\Delta t  \to 0} \sum_{n=0}^{N-1} h\left(\frac{X(t_{n+1}) + X(t_n)}{2}\right) \int_{t_0+n \Delta t}^{t_0+(n+1)\Delta t} \xi(t')dt'\,,
\end{split}
\end{align}
(which is somewhat different from choosing $\alpha=0$ in \eqref{E:defSI} above.)
Often, one associates a stochastic term coming from physical processes with the Stratonovich integral \eqref{E:defStra}. 
In order to distinguish the two prescriptions we will use 
\begin{subequations}
\label{E:ringnotation}
\begin{align}
\label{E:Ito}
	\frac{\partial}{\partial t} X(t) &= g(X(t)) + h(X(t))\xi(t) & &\hbox{It\^{o}} \,, \\
\label{E:Stratonovich}
	\frac{\partial}{\partial t} X(t) &= g(X(t)) + h(X(t)) \circ \xi(t)  & &\hbox{Stratonovich} \,,
\end{align}
\end{subequations}
in the remainder of this appendix. In the rest of this work we use the Stratonovich scheme.

In many instances, the stochastic term $\xi(t)$ in \eqref{E:ringnotation} is associated with white noise (with zero mean) where
\begin{equation}
\label{E:whitenoise}
	\overline{\xi(t)} = 0\,,
	\qquad
	\overline{\xi(t')\xi(t)} = D \delta (t-t')\,,
\end{equation}
with $D>0$,\footnote{
Recall that the spectral density $S(\omega)$ is the Fourier transform of the autocorrelation function $R(t-t') = \overline{\xi(t)\xi(t')} = \int S(\omega) e^{i \omega (t-t')}d\omega$. (Assuming $\overline{\xi(t)} = 0$.) Thus, $\overline{\xi(t)^2} = \int S(\omega) d\omega$ and $S(\omega)d\omega$ specifies the contribution of frequencies in the range $(\omega,\,\omega+d\omega)$ to the variance of $\xi(t)$. The power spectrum for white noise is uniform.} and higher moments of $\xi$ are such that
\begin{equation}
\label{E:Gaussian}
	\int_t^{t+T} \xi(t')dt' = \sqrt{D T} N(0,1)\,, 
\end{equation}
with $N(\mu,\sigma)$ denoting a normal distribution with average $\mu$ and standard deviation $\sigma$. In this case, one can relate the It\^{o} and Stratonovich differential equations 
via
\begin{equation}
\label{E:SToI}
	g + h \circ \xi = \left(g + \frac{D}{2} h' h \right) + h \xi \,.
\end{equation}
To obtain \eqref{E:SToI} we made use of
\begin{align}
	\int_t^{t+\Delta t}  h(X(t')) \circ \xi(t') dt' &= h\left(\frac{X(t) + X(t+\Delta t)}{2}\right)\int_t^{t+\Delta t} \xi(t')dt'\,, \\
	\int_t^{t+\Delta t}  h(X(t'))  \xi(t') dt' &= h\left(X(t)  \right)\int_t^{t+\Delta t} \xi(t')dt'\,,
\end{align}
equations \eqref{E:Gaussian} and \eqref{E:ringnotation} and have expanded all expressions to $\mathcal{O}\left((\Delta t)^\frac{3}{2}\right)$.

One somewhat unusual feature of the It\^{o} integration scheme is that the usual rules of calculus are modified. For instance, for a smooth function $f$, with
\begin{equation}
	f(X(t+\Delta t)) = f(X(t)) + f'(X(t)) (X(t+\Delta t)-X(t)) + \frac{1}{2}f''(X(t))  (X(t+\Delta t)-X(t))^2 + \mathcal{O}\left( (\Delta t)^{\frac{3}{2}} \right) \,,
\end{equation}
we obtain
\begin{equation}
	df(X) =f'(X) \left(g(X)dt + h(X) \xi dt\right)  + \frac{D }{2} f''(X) h(X)^2 dt  
\end{equation}
where we made use of \eqref{E:Gaussian}. In contrast, under the Stratonovich scheme,
\begin{equation}
	f(X(t+\Delta t)) = f(X(t)) + f'\left(\frac{X(t+\Delta t) +X(t)}{2}\right)  (X(t+\Delta t)-X(t)) + \ldots \,,
\end{equation}
and the usual rules of calculus hold,
\begin{equation}
	df(X) = f'(X)  \left(g(X)dt + h(X) \circ \xi dt\right) \,.
\end{equation}

With an integral expression for $X(t)$ one can derive a discrete time evolution scheme by which one may evolve $X(t)$ numerically from some initial time $t_0$. Expanding the It\^o integral to linear order in $\Delta t$ we find the Euler-Maruyama approximation
\begin{equation}
\label{E:EMexpansion}
	X(t+\Delta t) = X(t) + h(X(t)) \sqrt{\Delta t D} \phi + g(X(t))\Delta t  \,,
\end{equation}
where $\phi$ is drawn from $N(0,1)$. A Stratonovich equation can be expanded in a similar manner by using \eqref{E:SToI}. When computing averages, the Euler Maruyama scheme provides a good approximation to $X$ up to order $\Delta t$. But when considering higher moments of $X$, the Euler-Maruyama scheme is good to order $\sqrt{\Delta t}$.
As in standard differential schemes there exist higher order numerical methods which improve convergence in $\Delta t$. For instance, the Milstein scheme is defined such that
\begin{equation}
	X(t+\Delta t) = X(t) + h(X(t)) \sqrt{\Delta t D} \phi + g(X(t))\Delta t  + \frac{D}{2} h(X(t))h'(X(t)) \left(\phi^2 -1 \right) \Delta t \,,
\end{equation}
whose error is of order $\Delta t$, a factor of $\sqrt{\Delta t}$ higher than the Euler-Maruyama scheme.

So far we have considered ordinary SDE's. Our entire discussion can be uplifted to a set of coupled ordinary SDE's,
\begin{equation}
\label{E:coupledIto}
	\frac{\partial}{\partial t} \vec{X} = \vec{G}(\vec{X}) + H (\vec{X}) \vec{\xi}
\end{equation}
with $\vec{X}$ and $\vec{G}$, $n$ dimensional vectors, and $H$ an $n\times n$ matrix. In this work we will specialize to interactions of the form
\begin{equation}
\label{E:OurH}
	H_{ij}(\vec{X}) = h_i(\vec{X}) \delta_{jn}\,,
\end{equation}
such that a single stochastic source, $\xi$, is present.
The Stratonovich version of \eqref{E:coupledIto} with \eqref{E:OurH} can be recast in It\^{o} form via an analog of \eqref{E:SToI}
\begin{equation}
\label{E:SToIvector}
	\vec{G} + \vec{h} \circ \xi = \left(\vec{G} + \frac{D}{2} \left(\vec{h} \cdot \vec{\nabla}_X\right) \vec{h} \right) + \vec{h} \xi\,.
\end{equation}
The Milstein scheme for solving the It\^{o} SDE \eqref{E:coupledIto} now reads
\begin{equation}
\label{E:OurMilstein}
	\vec{X}(t+\Delta t) = \vec{X}(t) + \vec{h}(t) \sqrt{D\Delta t} \phi +\vec{G}(t) \Delta t + \frac{D}{2}  \left(\vec{h} \cdot \vec{\nabla}_{X} \right) \vec{h} (\phi^2-1)\Delta t\,.
\end{equation}

\end{appendix}	
	
\bibliographystyle{plain}
\bibliography{Tbib}

\end{document}